# Multiple quantum wells for passive ultra short laser pulse generation

**R. Quintero-Torres**[*1], **E. Vázquez-Cerón**[2], **E. Rodríguez-Rodríguez**[2], **Andreas Stintz**[3], and **Jean-Claude Diels**[3]

[1] Centro de Física Aplicada y Tecnología Avanzada, UNAM Apartado Postal 1-1010 76000 Querétaro, México

[2] Departamento de Electrónica, Universidad Autónoma Metropolitana-Azcapotzalco, Av San Pablo 180, México D.F. 02200

[3] Department of Physics and Astronomy, CHTM, University of New Mexico, Albuquerque NM 87131, USA

[*] Corresponding author: e-mail: rquintero@fata.unam.mx, Phone: +52 (55) 56234166, Fax: +52 (55) 56234165

**PACS** 42.60.Fc, 78.67.De

Solid state lasers are demanding independent control in the gain media and cavity loss to achieve ultra short laser pulses using passive mode-locking. Recently, laser mode-locking is achieved with a MBE structure with multiple quantum wells, designed to achieve two functions; Bragg mirror and changes in absorption to control the cavity dynamics. The use of an AlGaAs/AlAs Bragg mirror with a 15 nm GaAs saturable absorber used in a Cr:LiSAF tuneable laser proved to be effective to produce femtosecond pulses. The use of saturable absorbers thus far is a trial and error procedure that is changing due to the correlation with more predictive procedures.

## 1 Introduction

The solid state laser is evolving toward new level of control in the cavity dynamics including pulse time duration and acceptance of perturbations, particularly a versatile control in the gain media and cavity loss is demanded to achieve short laser pulses. This is particularly valid for ultra short pulses for the reason that the active mode-locking limits the pulse duration by few picoseconds.

Passive mode-locking is used to obtain sub picoseconds pulses with the help of two main effects, control in the nonlinear refractive index (Kerr effect) [1,2] and control in the nonlinear absorption [3,4]. The use of nonlinear absorption was originally included in the laser cavity in the form of a liquid with saturable absorption. Recently, laser mode-locking is achieved with a MBE structure with multiple quantum wells, designed to achieve two functions; Bragg mirror and absorption saturation as a result of the increasing intracavity power.

With a single transversal mode in a laser cavity, several longitudinal modes are possible to exist as long as the gain is larger than the loss and they are defined by the wavelength of operation and the cavity length. The control over the phase for all the longitudinal modes can be accomplished by two generic means; active mode-locking where an optical modulator in the cavity is controlled externally introducing periodic losses and passive mode locking where the phase control is achieved with a nonlinear optical effect, refractive or absorptive, modulated by the same intensity in the laser. With phase control over the longitudinal modes the radiation change from random intensity typical of thermal fluctuations to periodic pulsed intensity where the pulse duration is inversely proportional to its spectral width for transform limited pulses

## 2 Cavity long time dynamics

The sequence to produce a pulse assisted by the non linear absorption is as follows:
1) The laser does not work; the loss is bigger than the gain, the amplification is linear and the absorption is linear. 2) The laser works, however the operation is not pulsed. The gain surpasses the threshold, and the noise amplification favors the modes with larger gain. 3) The laser is working, and the pulse begins to take shape. The amplification is linear and the absorption is nonlinear due to the increase in the peak intensity. The absorbent material becomes transparent for some fluctuations and the loss increase for small fluctuations. The shape is defined because the intensity increases in the regions of larger intensity and the intensity decrease in the small intensity region. 4) Nonlinear amplification takes place, the pulse is passing several times by the gain media and a repetitive train of pulses appear at the output of the laser. 5) It reaches a steady state or the gain is depleted and the laser is extinguished.

In the longer time scale the recovery time of the gain media and the absorber are responsible for the continuous-wave mode-locking condition CWML, constant envelope for the mode-locked pulses or the Q-switching mode-locking condition QSML, Q-switched envelope for the mode-locked pulses.

To describe the average power in the laser, the pulse shape is unimportant because the power change is in a time scale larger than the pulse duration ($\tau$) or the cavity round trip time ($T$). The dynamics of the average power, that is proportional to the pulse energy ($W$), is defined by the gain factor after each round trip ($g$) and the loss factor in the saturable absorber after each round trip ($q$). The rate equations for the gain factor and the pulse energy are:

$$\frac{dW}{dt} = \frac{g-l-q}{T}W \quad \text{and} \quad \frac{dg}{dt} = -\frac{g-g_0}{\tau_L} - \frac{W}{W_L \tau_L}g \tag{1}$$

Where $l$ is the linear loss factor at each round trip, $g_0$ is the gain factor without lasing, proportional to the pump flux and the mode overlapping between pump and laser. The gain media is characterised by the time constant ($\tau_L$) and the effective saturation energy ($W_L$). The loss factor in the saturable absorber should include the photobleaching as well as the inverse saturable absorption; two photon absorption and free carriers. A physically intuitive model [5] for the absorber with two parameter that define the bleachable absorption ($a$) and the energy to start considering the inverse saturable absorption ($W_0$) is next,

$$q = a(W - W_0)^2 \tag{2}$$

In this condition the laser starts if $g_0 = l + aW_0^2 \equiv g_0^A$, the CWML condition is achieved if $W$ and $g$ are constant with time, and it is obtained if $g_0 = l\left(1 + \frac{W_0}{W_L}\right) \equiv g_0^B$ is reached or surpassed. If $W_0 > W_L$ as it is usually the case, then in between $g_0^B$ and $g_0^A$ the QSML condition is observed. A typical set of parameters is; $T$=10 ns, $l$=30%, $\tau_L$=2 µs, $W_L$=13 nJ, $a$=8x10$^{-5}$ nJ$^{-2}$, $W_0$=25 nJ; producing $g_0^A = 35\%$ and $g_0^B = 87\%$.

### 3 Bragg mirror and saturable absorber
The design of a Brag mirror is a standard procedure in vertical cavity surface emitting laser VCSEL, for this range of wavelengths AlAs/AlGaAs is typical, working with the thickness and the refractive index of the layers a very good mirrors are obtained (see Figs. 1 and 2). The absorber is 15 nm low life-time carriers with 2% aluminium between 10% aluminium walls. The behaviour of the absorber is defined by the quantum well configuration; the transition takes place between subbands of the same order in the valence and conduction bands. At the origin of the k-space the subbands for the holes are no longer degenerated.

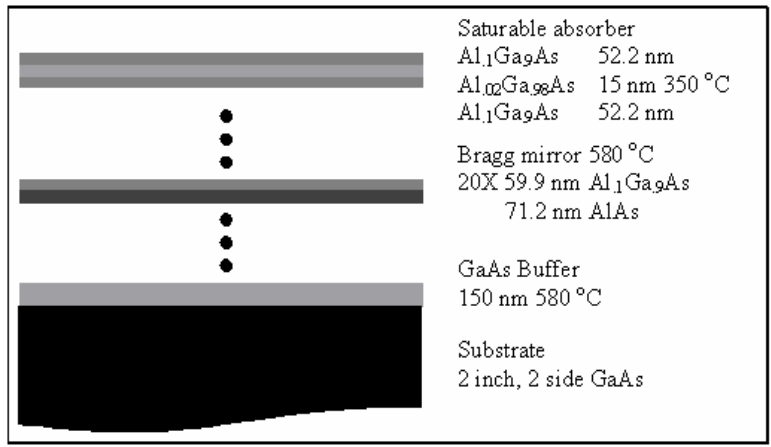

**Fig. 1** MQW structure, mirror design to work around 850 nm and absorber at 850 nm, labelled IV in figure 3.

The qualitative ideas for the response of the absorber in the photobleaching region are related with the photons carried by the laser energy $W\lambda/\hbar c$; and the density of states in the quantum well, times the volume defined by the laser. Beyond that region the nonlinear absorption due to two photons and the free carriers take control

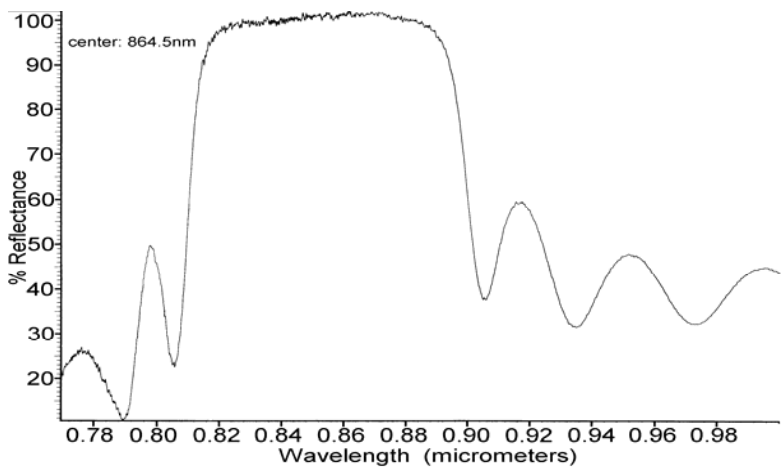

**Fig. 2** Reflectance at low intensity for the MQW of figure 1.

Using the results for the first transition from the heavy holes level to the electron level [6] is possible to conclude that changing the QW from 20 to 15 nm increases the first energy transition by 5 meV. Increasing the aluminium content from 0% to 2% in the well increases the first transition in 25 meV. Increasing the Al content in the walls from 10% to 30% increases the first energy transition in 5 meV.

## 4 LiSAF laser

The use of AlGaAs/AlAs Bragg mirror with a 15 nm GaAs saturable absorber used in a Cr:LiSAF tuneable laser proved to be effective to produce femtosecond pulses. The aim in developing laser sources more versatile is to follow the steps of evolution of electronics, and consider laser as the system rather than an independent unit, a light source. To build a solution requires introducing elements in the laser cavity and the desired and prevented effects are multiplied and can drift the laser out of control. Solid state saturable absorbers in addition to help in the old aim of improving the performance of classic applications of lasers are opening the possibility of introducing real system control.

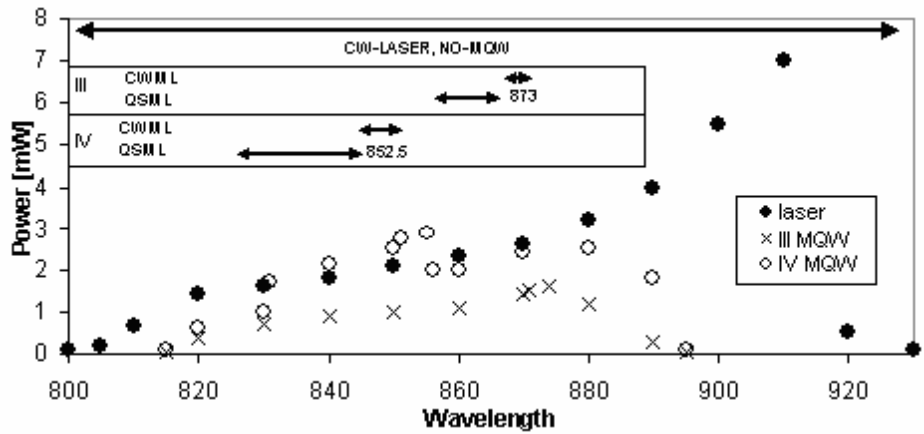

**Fig. 3**  Behaviour of a LiSAF laser without MQW, and with two designs of MQW.

$Cr^{3+}$:$LiSrAlF_6$ (LiSAF) belongs to the family of tuneable lasers using lattice vibrations, the tuneability range covers from 780-920 nm. Similar to Ti:Sapphire but with the advantage to be optically pumped with AlGaInP laser at 670 nm. And with a shortest pulse time duration reported of 10 fs[7].

Figure 3 presents with full circles the tuneability imposed by optical elements used, we did not use an output coupler, the radiation is measured behind a good mirror and the radiation in such case is not pulsed. Changing one mirror for a MQW (labelled III, cross mark) allows CWML to exist in a small range near 873 nm, with the laser pump used, such MQW is different from figure 1 only in the well material 0% aluminium. The open circles are produced with the MQW (labelled IV).

**Acknowledgements** We gratefully acknowledge the help of Thomas Rotter of CHTM.

### References
[1] D. E. Spence, P. N. Kean, W. Sibbett, Opt. Lett. **16**, 42 (1991).
[2] F. Salin, J. Squier, M. Piché, Opt. Lett. **16**, 1674 (991).
[3] U. Keller et al., IEEE J. Select. Topics Quantum Electron. **2**, 435 (1996).
[4] T. R. Schibli, E. R. Thoen, F. X. Kartner, E. P. Ippen, Appl. Phys. B (2000).
[5] Thien Trang Dang, A. Stintz, J.-C. Diels, Y. Zhang, Active Solid State Short Pulse Laser Gyroscope, ION 57th Annual Meeting - Session D4, Albuquerque, New Mexico (June 2001).
[6] P. Pereyra, Phys. Rev. Lett. **80**, 2677 (1998).
[7] S. Uemura and K. Torizuka. IEEE J. Quantum Electron. **39**(1), 68 (2003).